\documentclass[aps,pra,twocolumn,superscriptaddress,showpacs,amsmath,amssymb,amsfonts]{revtex4-1}
\usepackage{bm}
\usepackage{graphicx}
\usepackage{subfigure}
\usepackage{array}
\usepackage[usenames,dvipsnames]{color}
\usepackage{natbib}
\usepackage{pifont}
\usepackage[breaklinks=true,pdfborder={0 0 0},colorlinks=true,linkcolor=MidnightBlue,citecolor=MidnightBlue,urlcolor=MidnightBlue]{hyperref}
\newcommand{\mbf}[1]{{\textbf #1}}
\newcommand{\redcircle}[0]{{\color{red} \scriptsize \ding{108}}}
\newcommand{\greensquare}[0]{{\color{green} \scriptsize \ding{110}}}
\newcommand{\bluetriangle}[0]{{\color{blue} \scriptsize \ding{115}}}

\begin{document}
\title{Spontaneous Crystallization And Filamentation Of Solitons In Dipolar Condensates}

\author{Kazimierz {\L}akomy}
\affiliation{Institut f\"ur Theoretische Physik , Leibniz Universit\"at, Hannover, Appelstrasse 2, D-30167, Hannover, Germany}
\author{Rejish Nath}
\affiliation{Max Planck Institute for the Physics of Complex Systems, N{\"o}thnitzer Strasse 38, D-01187 Dresden, Germany}
\affiliation{IQOQI and Institute for Theoretical Physics, University of Innsbruck, A-6020 Innsbruck, Austria}
\author{Luis Santos}
\affiliation{Institut f\"ur Theoretische Physik , Leibniz Universit\"at, Hannover, Appelstrasse 2, D-30167, Hannover, Germany}
\date{\today}

\begin{abstract}

Inter-site interactions play a crucial role in polar gases in optical lattices 
even in the absence of hopping. We show that due to these long-range interactions 
a destabilized stack of quasi-one dimensional Bose-Einstein condensates develops a 
correlated modulational instability in the non-overlapping sites. Interestingly, 
this density pattern may evolve spontaneously into soliton filaments or into a 
checkerboard soliton crystal that can be so created for the first time in 
ultra-cold gases. These self-assembled structures may be observed under realistic 
conditions within current experimental feasibilities.

\end{abstract}
\pacs{03.75.Lm, 03.75.Kk, 05.30.Jp}
\maketitle

\section{Introduction}
Recent experiments are opening new avenues for the study of the fascinating physics of 
dipolar gases \cite{Baranov:2008,Lahaye:2009}. These gases present a significant electric 
or magnetic dipole-dipole interactions, which being long-range and anisotropic differ 
significantly from the short-range isotropic interactions usually dominant in quantum 
gases. Ultra-cold polar gases in optical lattices are particularly interesting. Contrary 
to the non-dipolar case, polar lattice gases are characterized by significant non-local 
inter-site interactions that result in a rich variety of novel physical phenomena 
\cite{Lahaye:2009,Trefzger:2011}. Remarkably, the inter-site interactions play a crucial 
role even in deep lattices where hopping is negligible. In particular, dipolar Bose-Einstein 
condensates (BECs) in non-overlapping lattice sites share common excitations modes. This 
collective character enhances roton-like features in the excitation spectrum \cite{Klawunn:2009} 
and modifies the BEC stability, as recently shown experimentally \cite{Mueller:2011}.

Quasi-1D geometries allow for the existence of BEC solitons and hence 
modulational instability in these systems leads to the formation of 1D patterns, 
so-called soliton trains \cite{Strecker:2002}. On the contrary, dynamical instability in 
higher-dimensional BECs is typically followed by condensate collapse \cite{Donley:2001}. 
In consequence, solitons patterns in higher dimensions, as e.g. a 2D crystal of solitons, 
are fundamentally prevented in non-polar BECs.

In this paper we show that the destabilization of a dipolar BEC confined in a stack 
of non-overlapping quasi-1D tubes may be followed by the spontaneous self-assembly of 
stable soliton filaments or a 2D checkerboard crystal of solitons, providing a route for the first 
realization of self-sustained 2D arrangements of BEC solitons. This dynamical self-assembly 
stems from the correlated character of the corresponding modulational instability. While for 
non-dipolar condensates the instability in each lattice site would develop independently, 
the non-local dipolar interactions couple the non-overlapping BECs to form a density 
pattern shared among all sites. As we show, correlated modulational instability may be 
observable in current Chromium~\cite{Lahaye:2007} and Dysprosium~\cite{Lev:2011} experiments. 

The dynamically formed soliton filaments resemble dipolar chains of classical dipoles 
\cite{Teixeira:2000}, as well as chains predicted for polar molecules \cite{Wang:2006,Duhme:2010}. 
However, compared to the latter, soliton filamentation is expected to occur for smaller dipole 
moments due to the many-body character of each soliton. Remarkably, inverting the sign of the 
dipolar interactions results in the development of an anti-correlated density pattern which may 
be followed by the spontaneous formation of a stable crystal of solitons. This 2D checkerboard 
crystal resembles the Wigner-like crystal predicted for polar molecules \cite{Buchler:2007,Pupillo:2008}. 
However, contrary to the latter, it is dynamically formed and self-maintained by a non-trivial 
interplay between intra-tube attractive and inter-tube repulsive dipolar interactions. 

\section{Model}
\begin{figure}[t]
\includegraphics[scale=1.0]{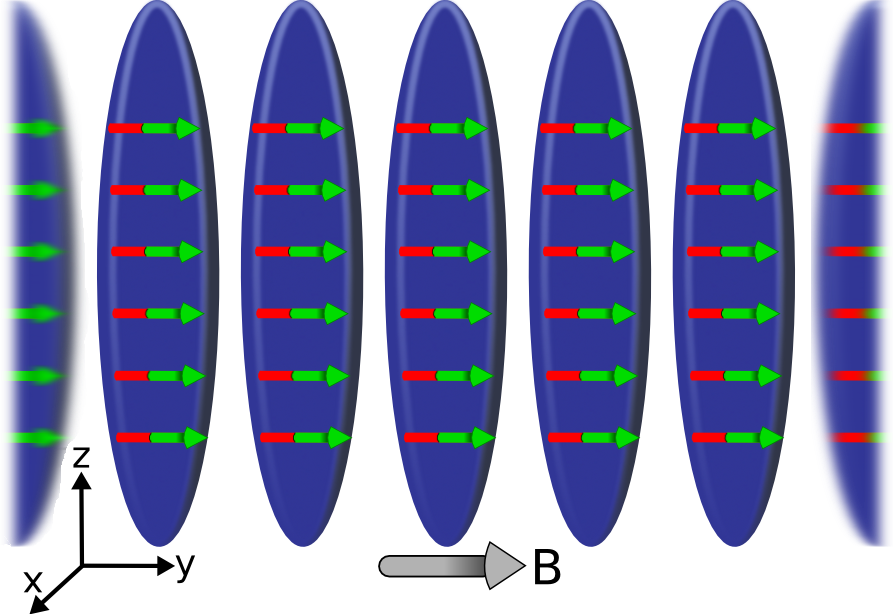}
\caption{(Color online) Scheme of the stack of disjoint quasi-1D dipolar BECs.}
\label{fig:setup}
\end{figure}

We study below a dipolar BEC confined in a stack of quasi-1D tubes formed by 
an optical lattice (Fig.~\ref{fig:setup}). The lattice is assumed to be sufficiently deep 
to suppress inter-site hopping. In each of the $N_{m}$ lattice sites the $xy$-confinement 
is approximated by a harmonic potential with frequency $\omega_{\scriptscriptstyle \perp}$, whereas 
for simplicity we assume no confinement along $z$ direction. We consider atoms with a magnetic 
dipole moment $\mu$ (the results are equally valid for electric dipoles, as e.g. polar molecules) 
oriented along $y$ direction by an external magnetic field. The dipoles interact with each other 
via the dipole-dipole potential $V_{d}\left({\mbf{r}} - \mbf{r'}\right) = g_{d} \left( 1 - 3 \cos^{2} 
\theta \right)/ \left| \mbf{r} - \mbf{r'}\right|^{3}$, where $g_d= \mu_{0} \mu^{2}/4 \pi$, 
with $\mu_{0}$ being the vacuum permeability and $\theta$ the angle formed by the vector 
joining the two interacting particles and the dipole moment direction. 

%%%%%%%%%%%%%%%%%%%%%  Fig.1.  %%%%%%%%%%%%%%%%%%%%%%
%%%%%%%%%%%%%%%%%%%%%%%%%%%%%%%%%%%%%%%%%%%%%%%%%%%

We assume the chemical potential much smaller than $\hbar \omega_{\scriptscriptstyle \perp}$ 
(this assumption is self-consistently verified in our calculations). Hence, we can 
factorize the BEC wave function at each site $j$, $\Psi_{j} \left( {\mbf{r}} \right) = 
\phi_{j} \left( x,y \right) \psi_{j} \left( z \right)$, with $\phi_{j}\left( x,y \right)$ 
the ground-state wave function of the $xy$ harmonic oscillator. Treating the dipolar potential 
in the Fourier space \cite{Goral:2002} we arrive at a system of $N_m$ coupled 1D Gross-Pitaevskii 
equations describing the BEC stack:
\begin{multline}
 \imath \hbar \partial_{t} \psi_{j} \left( z \right) = \Biggl[ -\frac{\hbar^2}{2m} \partial_{z}^{2} 
 + \frac{g}{2 \pi l_{\scriptscriptstyle \perp}^{2}} \left| \psi_{j} \left( z \right) \right|^{2} \\ 
 + \frac{g_d}{3}\sum_{m=0}^{N_{m}-1} \int \frac{d k_z}{2 \pi} e^{\imath k_z z}  \hat{n}_{m} \left( k_z \right) 
 F_{mj} \left( k_z \right) \Biggr] \psi_{j} \left( z \right),
\label{eqn:1DGPsystem}
\end{multline}
where $\hat{n}_{m} \left( k_z \right)$ is the Fourier transform of the axial density $n_m(z)$ at site $m$, 
\begin{align}
F_{mj} \left( k_z \right) = \! \int & \!\frac{d k_x d k_y}{\pi} \, 
\left( \frac{3 k_{y}^{2}}{ k_{x}^{2} + k_{y}^{2} + k_{z}^{2} } - 1 \right) \nonumber \\
& \times e^{-\frac{1}{2} \left( k_x^2 + k_y^2 \right) l_{\scriptscriptstyle \perp}^{2} - \imath k_y \left( m - j \right) \Delta },
\end{align}
$l_{\scriptscriptstyle \perp} = \sqrt{\hbar/m \omega_{\scriptscriptstyle \perp}}$ is the $xy$ oscillator length, 
$\Delta$ is the lattice spacing, and $g=4 \pi a \hbar^2/m$. Note that for a fixed ratio 
$\Delta/l_{\scriptscriptstyle \perp}$ the physics of the system is governed by the values of $g$ and $g_d$.
%%%%%%%%%%%%%%%%%%%%%  Fig.2.  %%%%%%%%%%%%%%%%%%%%%%
\begin{figure}[b]
\includegraphics[width=0.40\textwidth]{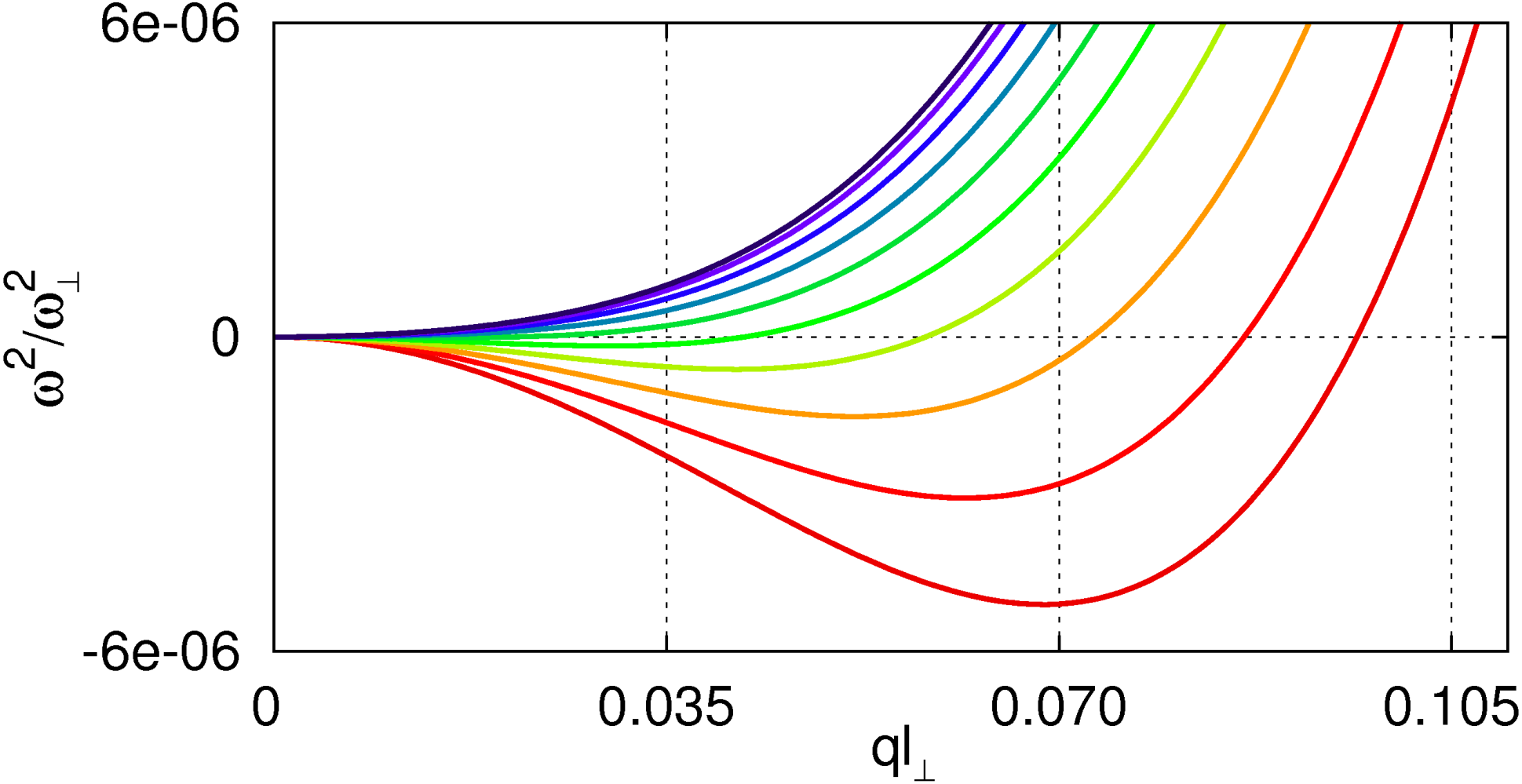}
\caption{(Color online) Bogoliubov spectrum for a $^{52}$Cr BEC ($\mu = 6 \mu_B$, 
where $\mu_B$ is the Bohr magneton) with a density $10^{14}$ cm$^{-3}$ and 
$a=-8.5 a_0$ ($a_{0}$ is the Bohr radius), occupying $N_m=10$ sites of a lattice 
with the inter-site spacing $\Delta = 512$ nm and a lattice depth of $13.3$E$_{R}$ 
(recoil energy), which results in the $\omega_{\scriptscriptstyle \perp}= 2 \pi \cdot 26.7$ kHz, 
and $l_{\scriptscriptstyle \perp}=85.3$ nm. Here, $q_{c}=0.07/l_{\scriptscriptstyle \perp}$.}
\label{fig:bogoliubovSpec}
\end{figure}
%%%%%%%%%%%%%%%%%%%%%%%%%%%%%%%%%%%%%%%%%%%%%%%%%%%

\section{Linear regime: Bogoliubov modes}
\label{sec:LinearRegime}
Starting from a homogeneous on-site linear density $n_0$ we are interested in the 
dynamics that follows the destabilization of the condensate after an abrupt change of 
the scattering length $a$. A substantial insight into the first stages of the post-instability 
dynamics is provided by the analysis of the elementary excitations of the condensate. To this 
end we introduce a perturbation of the homogeneous solution, $\psi_{j} \left( z, t\right) = 
\left[ \sqrt{n_{0}} + \chi_{j} \left( z, t \right) \right] e^{-\imath \mu_{j} t/ \hbar}$, with 
$\chi_{j} \left( z, t \right) = u_{j} e^{\imath \left( zq -\omega t\right)} + v^{*}_{j} 
e^{-\imath \left( zq -\omega t\right)}$, where $\mu_j$ is the chemical potential in a site 
$j$, and $q$ and $\omega$ are the $z$-momentum and the frequency of the elementary excitations, 
respectively. Employing this ansatz in Eq.\eqref{eqn:1DGPsystem} we arrive at the 
corresponding Bogoliubov-de Gennes equations yielding the excitation spectrum and the 
Bogoliubov coefficients $u_j$ and $v_j$. Interestingly, even in absence of hopping, 
dipolar inter-site interactions result in a collective character of the excitations that 
are shared by all sites. In consequence, 
the excitation spectrum acquires a band-like character \cite{Klawunn:2009} as depicted 
in Fig.~\ref{fig:bogoliubovSpec}. 
%%%%%%%%%%%%%%%%%%%%%  Fig.3.  %%%%%%%%%%%%%%%%%%%%%%
\begin{figure}[t]
\includegraphics[scale=1.2]{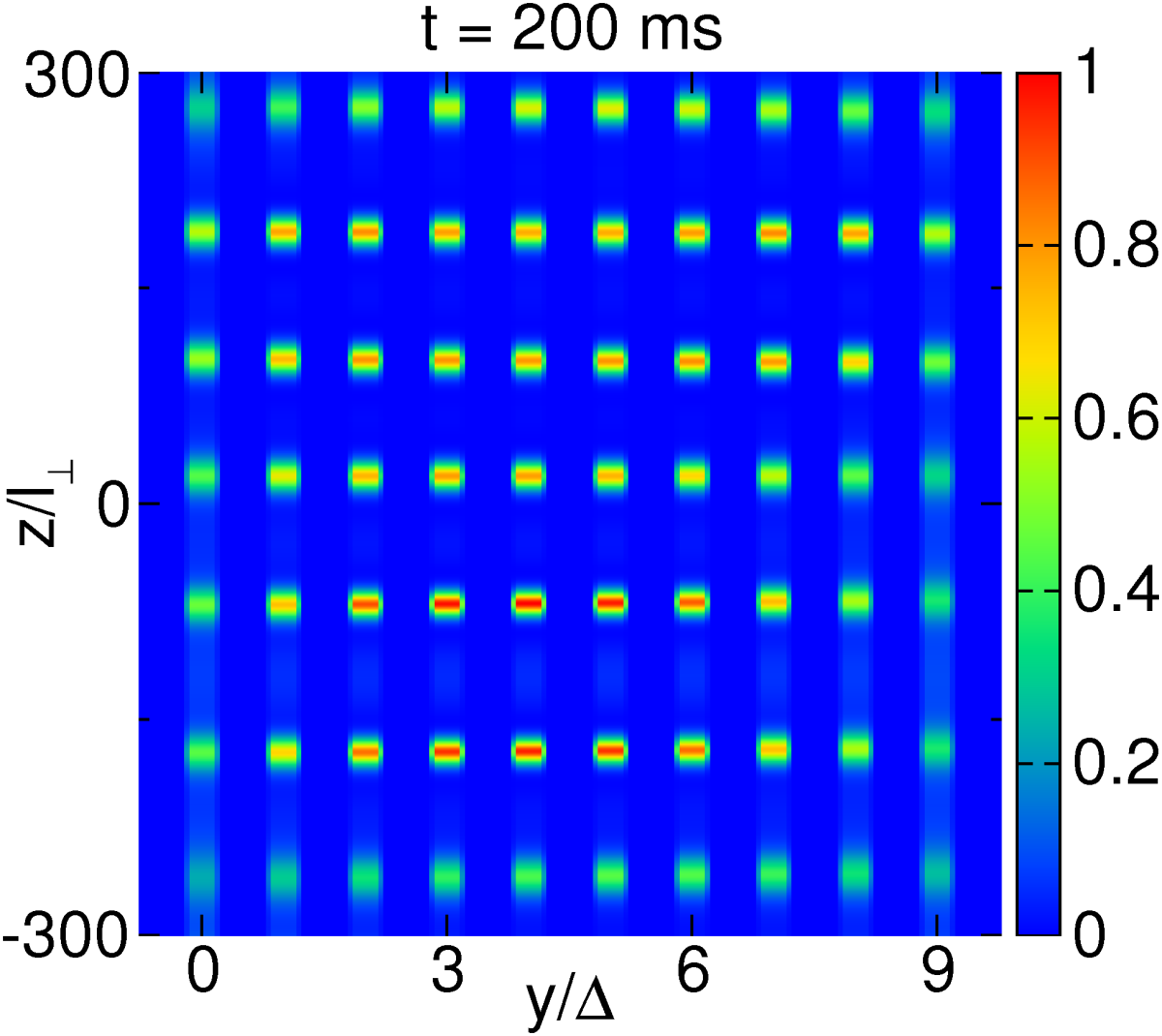}
\end{figure}
%%%%%%%%%%%%%%%%%%%%%%%%%%%%%%%%%%%%%%%%%%%%%%%%%%%
%%%%%%%%%%%%%%%%%%%%%  Fig.3.  %%%%%%%%%%%%%%%%%%%%%%
\begin{figure}[t]
\begin{center}
\vspace{-0.3cm}
\includegraphics[scale=1.0]{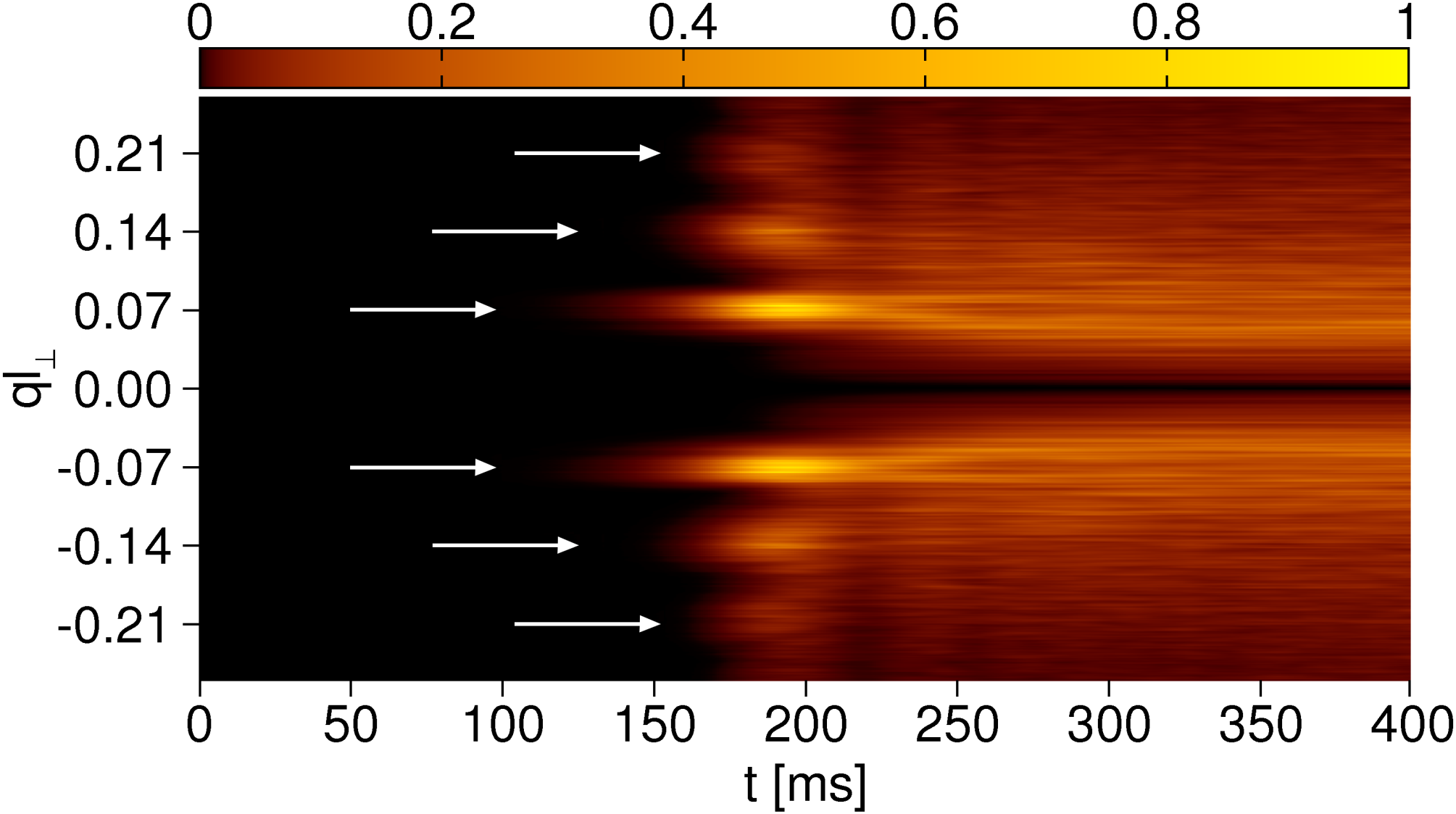}
\caption{
(Color online) (top) BEC wave function's density distribution after $200$ ms of time evolution 
for the same parameters as in Fig.~\ref{fig:bogoliubovSpec}. For plotting purposes the 
$y$-width of the tubes has been magnified. (bottom) Dynamics of the Fourier transform of 
the associated column density $\Sigma(z,t)$. The dominating $q=0$ peak has been removed for 
clarity and the remaining distribution has been normalized to the maximum. The arrows indicate 
the harmonics of $q_c$.}
\label{fig:FFTdynamics}
\end{center}
\end{figure}
%%%%%%%%%%%%%%%%%%%%%%%%%%%%%%%%%%%%%%%%%%%%%%%%%%%

Modes with imaginary frequency are associated with dynamical instability. For 
non-dipolar gases, inter-site interactions are negligible and hence all 
transverse modes remain degenerated. As a result, modulational instability 
develops independently in each site and no correlated density pattern occurs 
during the post-instability dynamics. The situation dramatically changes for sufficiently 
large dipole moment, as the inter-site interactions lift the degeneracy 
between the transverse modes. In particular, the most unstable mode becomes significantly 
more unstable than other modes, as shown in Fig.~\ref{fig:bogoliubovSpec}, governing 
the BEC dynamics within the linear regime. Crucially, this most unstable mode is not 
only characterized by a $z$-momentum $q_c$ (associated with the minimum of $\omega^2$ 
in Fig.~\ref{fig:bogoliubovSpec}) setting the modulational instability in each wire, but also 
by a transverse dependence along the $y$ direction locking the 
density pattern between sites. As a result, during the first stages of the post-instability 
dynamics a correlated modulational instability develops. Interestingly, our numerical simulations 
predict that this phenomenon may be observed in existing Chromium experiments \cite{Lahaye:2008} 
or even more pronouncedly with recently condensed Dysprosium atoms \cite{Lev:2011}.

Fig.~\ref{fig:FFTdynamics} (top) depicts the case of a $^{52}$Cr BEC destabilized
by an abrupt change of $a>0$ into a sufficient $a<0$ by means of a Feshbach resonance. 
The numerical solution of Eq.~\eqref{eqn:1DGPsystem} shows that despite the absence of 
inter-site hopping a correlated density pattern develops. As presented in 
Fig.~\ref{fig:FFTdynamics} (top) this instability pattern survives well into 
the non-linear regime where the density modulation cannot be considered any more as a 
perturbation of the original homogeneous on-site BECs. In typical experiments the density 
alignements may be more easily monitored investigating the column density 
$\Sigma(z) = \sum_{m} n_{m}(z)$. Contrary to the uncorrelated case, for which $\Sigma(z)$ 
would show no clear structure, the correlated instability results in periodically modulated 
$\Sigma(z)$. Fig.~\ref{fig:FFTdynamics} (bottom) shows the dynamics of the Fourier 
transform of $\Sigma(z,t)$ which is clearly characterized by the appearance of harmonics of
$q_c$ (compare Fig.~\ref{fig:bogoliubovSpec} and Fig.~\ref{fig:FFTdynamics} (bottom)).

\section{Filamentation}
\label{sec:Filamentation}
\begin{figure}
\vspace{-0.05cm}
\includegraphics[scale=1.21]{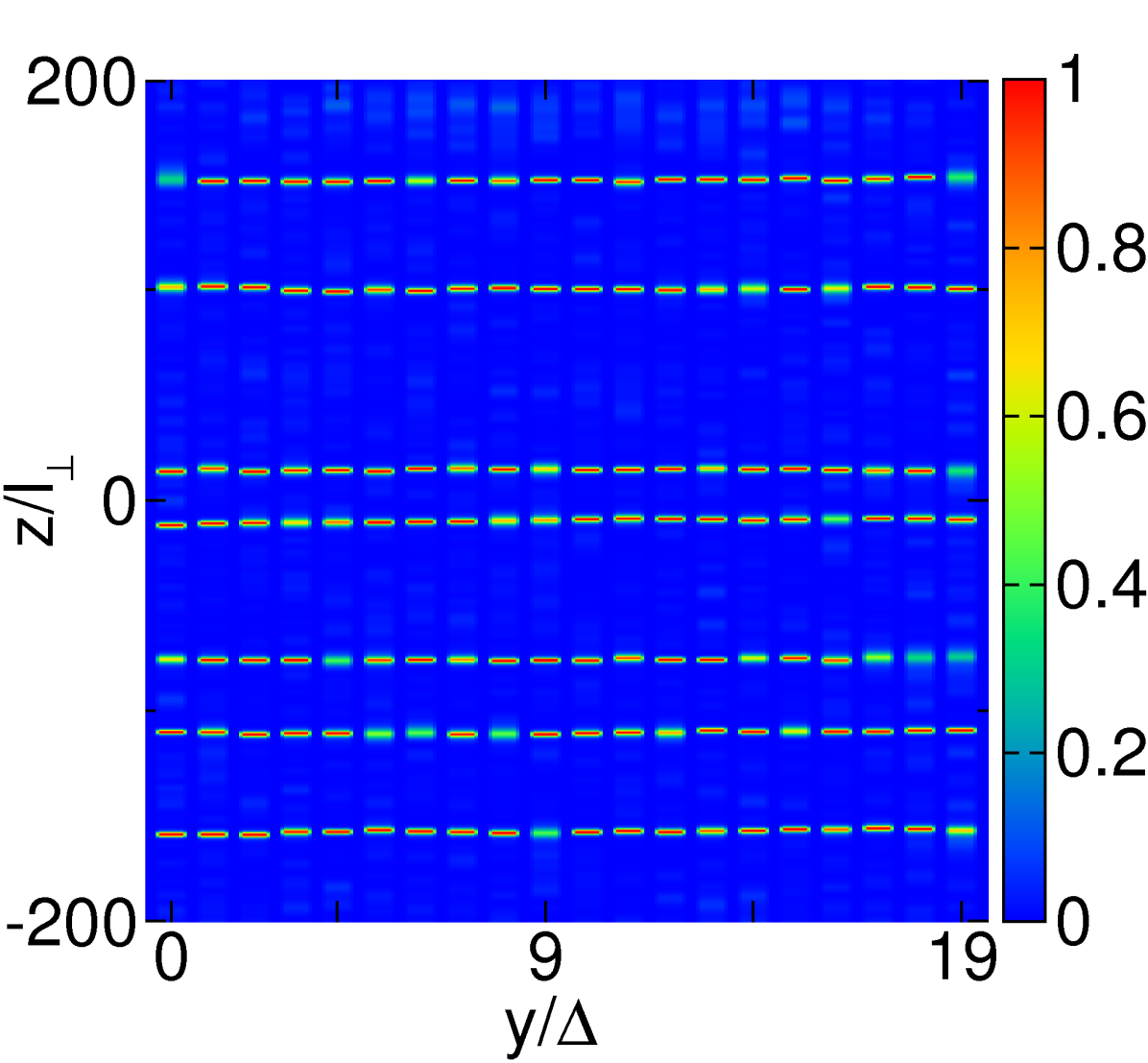}
\caption{(Color online) Filamentation of solitons. Here, a snapshot of time evolution of the 
BEC density distribution after $500$~ms for, in particular, $N_{m}\!\!=\! 20$ lattice sites, 
$\mu\!=\,$18$\mu_{B}$ and $\!a\!=\!-41.7a_{0}$. The remaining 
parameters are chosen as in the case of Fig.~\ref{fig:bogoliubovSpec}.}
\label{fig:filamRegimes}
\end{figure}

The density modulation depicted in Fig.~\ref{fig:FFTdynamics} (top) evolves into a 
correlated pattern of solitons. However, the solitons are created in an excited state, 
with both internal breathing excitation and center-of-mass motion. As a result, for 
insufficient dipolar interactions the correlated density modulation is destroyed during 
the subsequent non-linear time evolution. Consequently, the positions of solitons at 
different sites become uncorrelated, not differing qualitatively from the case of 
non-polar gases. Strong inter-site interactions crucially change this picture as the 
correlated solitons in neighboring sites experience an attractive inter-site potential. 
Approximating the solitons by Gaussians of width $\delta$, such that 
$l_{\scriptscriptstyle \perp}\ll \delta,\Delta$, the binding energy for two solitons 
acquires the form
\begin{equation}
E_b= (-2g_d/\Delta^3)G(\delta/\Delta),
\label{eqn:bindE}
\end{equation}
which differs from the binding energy between point-like solitons ($-2g_d/\Delta^3$) by the 
regularization function  
\begin{equation}
\!\!\!\!G(x)\!\simeq\! \frac{e^{1/4x^2}}{4\sqrt{2\pi} x^3}\!\left [\!
(x^2\!+\!1)K_{0}\!\!\left ( \!\frac{1}{4x^2}\! \right )\!+\!(x^2\!-\!1)K_{1}\!\!\left 
(\! \frac{1}{4x^2} \!\right )\! \right ]
\end{equation}
with $K_n$ the modified Bessel function of second kind. As a result of this 
inter-site soliton attraction, and although the initial periodicity of the modulation 
(as that of Fig.~\ref{fig:FFTdynamics} (top)) is generally lost, self-assembled soliton 
filaments form spontaneously (Fig.~\ref{fig:filamRegimes}) when the center-of-mass 
kinetic energy of the solitons acquired in the post-instability dynamics cannot 
overcome the binding energy given by Eq.~\eqref{eqn:bindE}. 

%%%%%%%%%%%%%%%%%%%%%  Fig.4.  %%%%%%%%%%%%%%%%%%%%%%
%%%%%%%%%%%%%%%%%%%%%%%%%%%%%%%%%%%%%%%%%%%%%%%%%%%

%%%%%%%%%%%%%%%%%%%%  Fig.5.  %%%%%%%%%%%%%%%%%%%%%%
\begin{figure}[t]
\includegraphics[width=1.0\columnwidth]{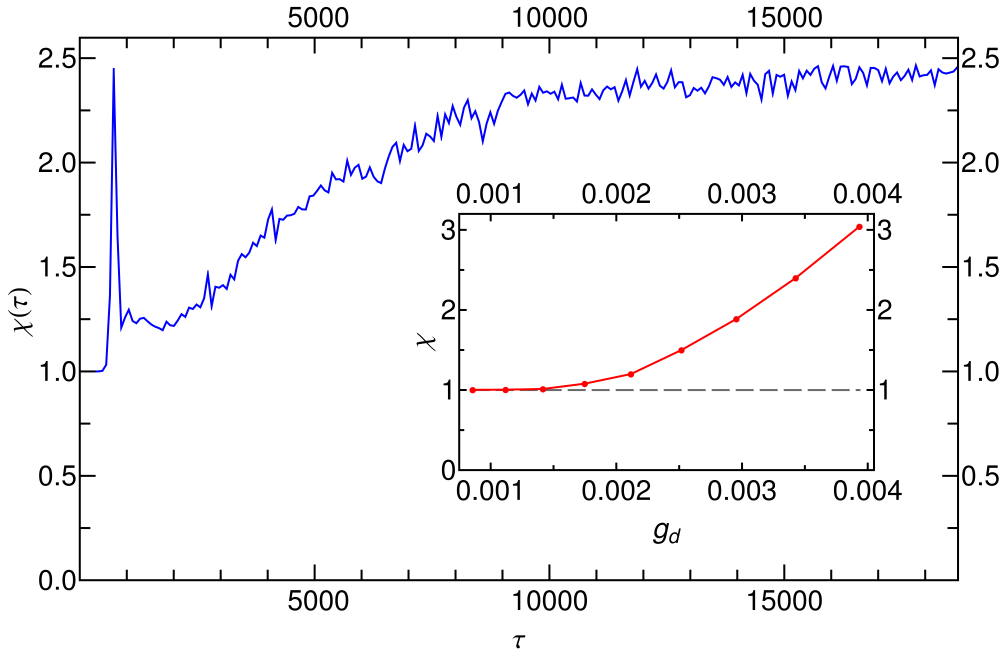} 
\vspace*{0.5cm} \\
\includegraphics[width=1.0\columnwidth]{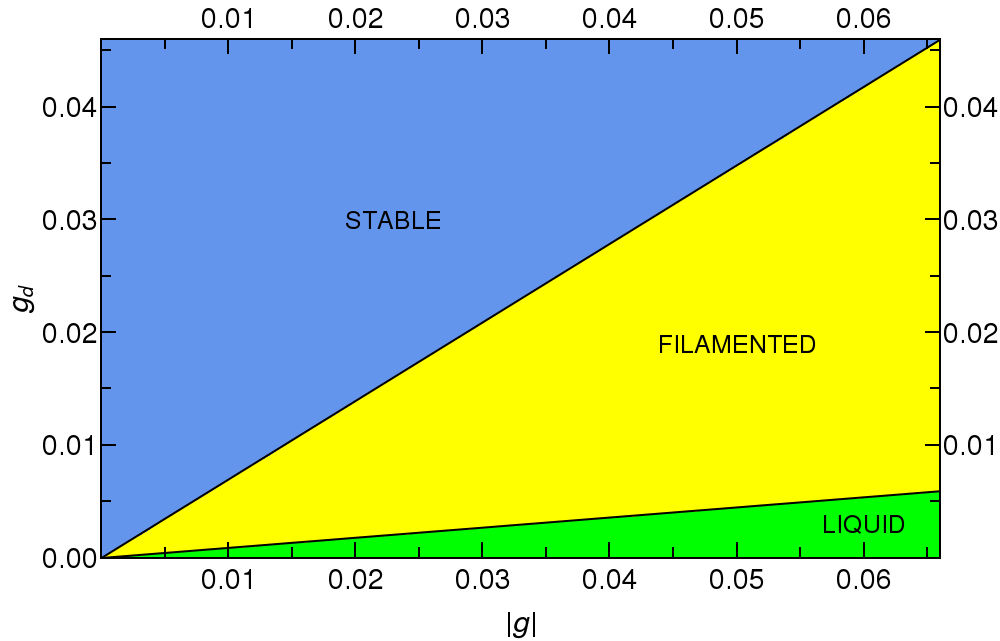}
\caption{(Color online) (top) Function $\chi(\tau)$ for a typical case within the 
filamentation regime~($g=-0.019$, $g_d=0.0034$, $\Delta=6 \,l_{\scriptscriptstyle \perp}$). 
In particular, for the parameters that we employed in Fig.~\ref{fig:bogoliubovSpec} the 
time ($t = \tau / \omega_{\scriptscriptstyle \perp}$) that we here consider equals $t=700$ ms. 
(inset) Time-averaged values of $\chi$ for long times, for different values of $g_d$ and 
constant $g=-0.019$. (bottom) Phase diagram of the possible regimes for $g<0$, $g_d>0$.} 
\label{fig:filam}
\end{figure}
%%%%%%%%%%%%%%%%%%%%%%%%%%%%%%%%%%%%%%%%%%%%%%%%%%%

In order to analyze the dynamical filamentation quantitatively we introduce at this point the 
time-dependent dimer correlation function for sites $m$ and $m'$
\begin{equation}
G_{m,m'}(z,t)=\int \! dz' \, n_m(z',t)n_{m'}(z'+z,t)
\end{equation}
and we define the normalized average dimer correlation $G^{\,n}(z,t)=G(z,t)/ \! \int \! dz \, G(z,t)$, 
with 
\begin{align}
G(z,t)=\frac{2}{N_{m}(N_{m}-1)}\sum_m \sum_{m'>m} G_{m,m'}(z,t).
\end{align}
A proper figure of merit describing the filamentation is provided by
\begin{equation}
\chi(t)=G^{\,n}(0,t)/{\bar G}^{\,n}(t), 
\end{equation}
where ${\bar G}^{\,n}(t)=\int \! dz \, G^{\,n}(z,t)^2$ is the mean value of $G^{\,n}(z,t)$. 
Such defined function $\chi(t)$ characterizes the tendency of the solitons at different 
sites to align into a filament. 

In the following analysis we consider a simplified case of three lattice sites. Fig.~\ref{fig:filam}~(top) 
shows $\chi(t)$ for a typical case within the filamentation regime (see discussion below). The sharp initial 
peak indicates the formation of the correlated density pattern shared among all sites at the initial stage of 
the time evolution, as discussed in section \ref{sec:LinearRegime}. Similarly to Fig.~\ref{fig:FFTdynamics}, 
also here the pattern is quickly destroyed as the system enters the non-linear regime. However, provided 
sufficently strong dipolar interactions, the inter-site soliton binding $E_{b}$ supports the formation of 
soliton filaments and in consequence $\chi(t)$ grows at larger times. Note that $\chi(t)$ eventually 
saturates remaining constant for times typically much longer than the usual experimental timescales. 

In contrast, no filamentation occurs if the dipolar coupling is insufficient. In this case, at long times 
$\chi(t)$ averages to $\chi=1$ indicating the absence of inter-site soliton-soliton correlation. Hence, 
driving the $g_d$ parameter from small to large values results in a transition from a non-filamented into 
a filamented configuration (see the inset of Fig.~\ref{fig:filam}~(top)). Ultimately, for a sufficiently 
large $g_d$ the repulsive on-site interactions compensate the attractive short-range interactions and the 
system remains stable. 

As a result, we distinguish three distinct regimes of dynamics in a stack of 1D dipolar gases: 
(i) unstable uncorrelated (soliton liquid), (ii) unstable filamented, and 
(iii) stable. As shown in Fig.~\ref{fig:filam}~(bottom), for a fixed value of $\Delta/l_{\scriptscriptstyle \perp}$ 
these regimes are determined by the ratio $g_d/|g|$. For the considered case of three sites and 
$\Delta/l_{\scriptscriptstyle \perp}=6$ the stability boundary line is given by $g_d/|g| = 0.70$, 
whereas the boundary line between the filamented and unstable non-filamentated regimes occurs at 
$g_d/|g| = 0.09$. For the case of $^{52}$Cr ($\mu = 6 \mu_B$) 
the filamentation occurs for $5.2<|a|/a_0<40.2$, whereas for 
$^{164}$Dy ($\mu = 10 \mu_B$) it occurs for $47.3<|a|/a_0<367.0$.

We note that for a larger number of sites the system is more unstable due to the inter-site 
attractive interactions~\cite{Klawunn:2009}. Also, the boundary between filamented and unstable 
non-filamented regimes is shifted towards larger $g_d$ values due to the enhanced 
role of the string-like modes of the filaments. Hence, even though the qualitative 
results will not be affected, increasing the number of sites will in general reduce the 
filamentation regime.

\section{Checkerboard soliton crystal}

\begin{figure}[t]
\includegraphics[scale=1.2]{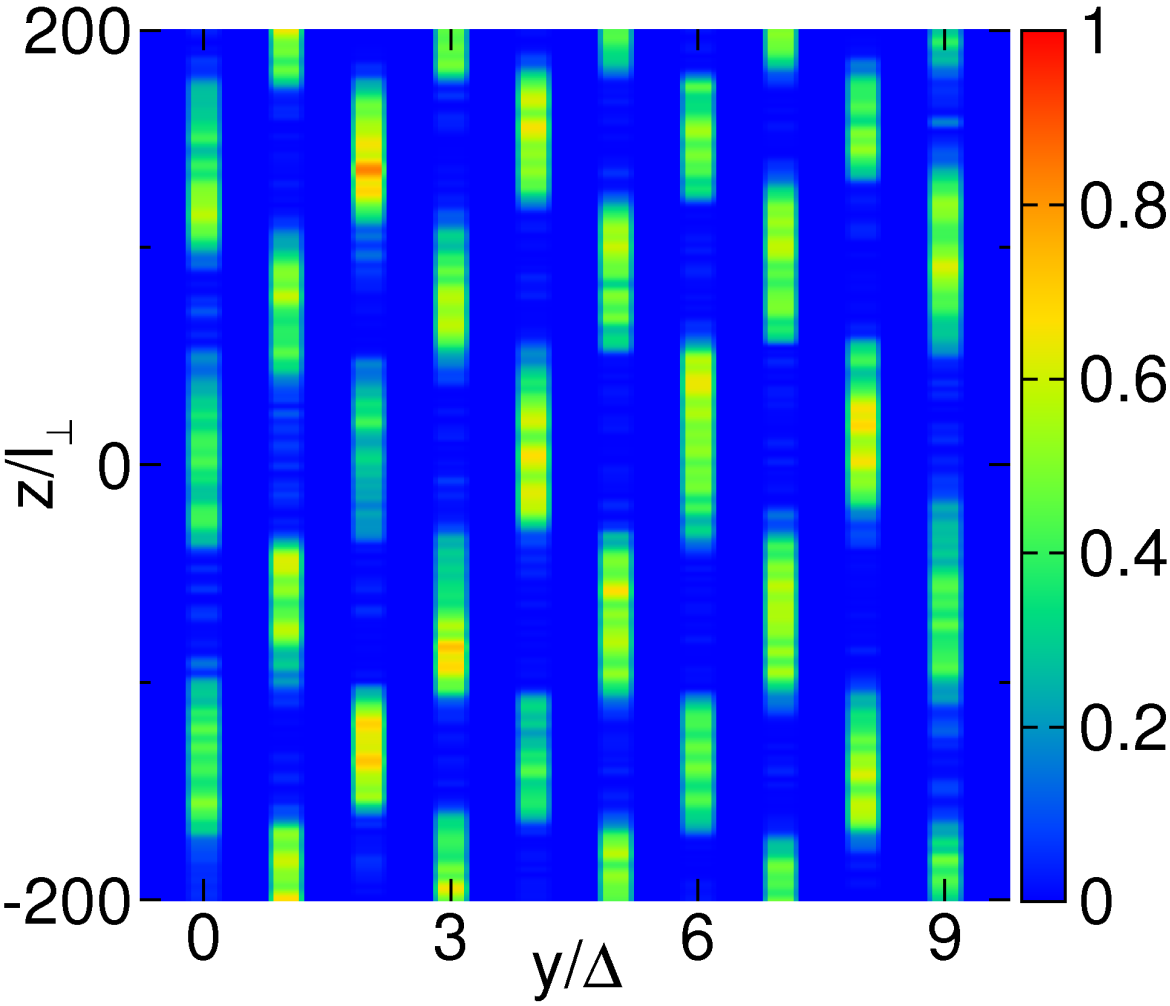}
\caption{(Color online) Spontaneous crystallization of solitons in the 
case of negative $g_{d}$. Here, $\!a\!=\!306a_0$, $\!\mu\!=\!36\mu_{B}$ 
and the remaining parameters such as in Fig.~\ref{fig:bogoliubovSpec}.}
\label{fig:crystalframes}
\end{figure}

Interestingly, the sign of $g_d$ may be inverted by means of transverse magnetic fields 
\cite{Giovanazzi:2002} or microwave dressing in the case of polar molecules \cite{Micheli:2007}. 
Note that, although we consider this case for its theoretical simplicity, qualitatively the same 
results may be obtained orienting the dipoles along the tubes. In both of these cases the emerging 
instability is characterized by the most unstable Bogoliubov mode presenting a staggered $y$-dependence 
that results in an anti-correlated density pattern with maxima in a given site aligned with minima in 
the neighboring ones. Strikingly, for a sufficiently strong dipole moment, this anti-correlated structure 
formed at the initial stage of the post-instability dynamics seeds the formation of a permanent checkerboard
soliton crystal in the non-linear regime, as shown in Fig.~\ref{fig:crystalframes}. 

Remarkably, while purely repulsive interactions sustain 2D Wigner-like 
crystals proposed for polar molecules \cite{Buchler:2007,Pupillo:2008}, 
the crystal of solitons is self-maintained by a subtle interplay 
of dipolar inter-tube repulsion and intra-tube attraction. Due to the 
anti-correlated character of the density modulation, solitons in neighboring 
sites provide an effective potential barrier that prevents mutually attracting 
solitons in the same tube to come together, hence keeping 
the crystal stable.

%%%%%%%%%%%%%%%%%%%%  Fig.6.  %%%%%%%%%%%%%%%%%%%%%%
%%%%%%%%%%%%%%%%%%%%%%%%%%%%%%%%%%%%%%%%%%%%%%%%%%%

%%%%%%%%%%%%%%%%%%%%  Fig.7.  %%%%%%%%%%%%%%%%%%%%%%
\begin{figure}[b]
\includegraphics[width=1.0\columnwidth]{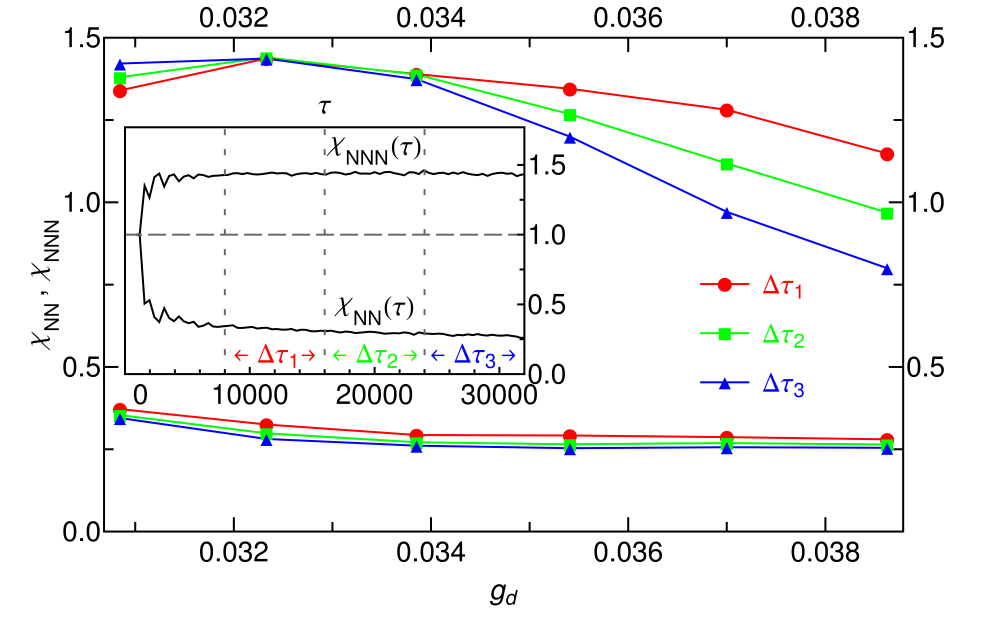} 
\vspace*{0.1cm} \\
\includegraphics[width=1.0\columnwidth]{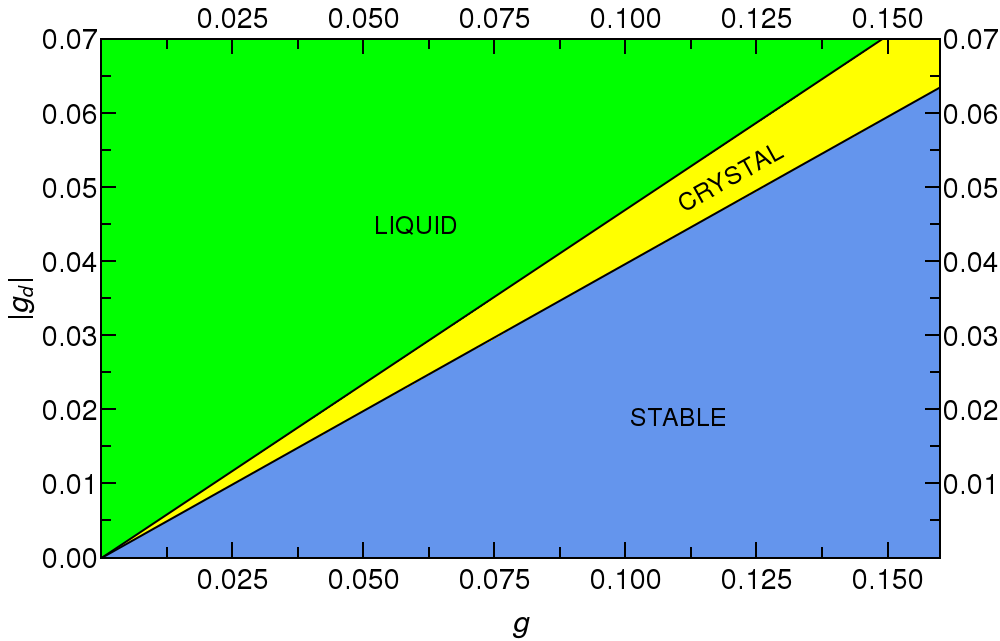}
\caption{(Color online) (top) The inset shows a typical example of time evolution of 
$\chi_{\scriptscriptstyle \,\!N\!N}(\tau)$ and $\chi_{\scriptscriptstyle \,\!N\!N\!N}(\tau)$ 
within the soliton crystal regime ($g=0.069$, $gd=-0.032$ and $\Delta=6 \,l_{\scriptscriptstyle \perp}$). 
We average $\chi_{\alpha}(\tau)$ within three different time intervals $\Delta \tau_{1,2,3}$, and we depict 
in the figure the corresponding averaged $\chi_{\alpha}$ (\redcircle, \greensquare, \bluetriangle) 
for different values of $g_d$ and constant $g=0.069$. Note that for $|g_d|/g>0.47$, 
$\chi_{\scriptscriptstyle \,\!N\!N\!N}$ decreases in time, indicating destruction of the 
checkerboard crystal. In particular, for the parameters we employed in Fig.~\ref{fig:bogoliubovSpec}, 
the time ($t = \tau / \omega_{\scriptscriptstyle \perp}$) that we here consider equals $t=1200$ ms. 
(bottom) Phase diagram of the possible regimes for $g>0$, $g_d<0$.}
\label{fig:checkerboard}
\end{figure}
%%%%%%%%%%%%%%%%%%%%%%%%%%%%%%%%%%%%%%%%%%%%%%%%%%%

In order to characterize quantitatively the dynamical formation of a soliton crystal, we employ 
the notation introduced in section \ref{sec:Filamentation}, defining the 
normalized averaged nearest-neighbor (NN) and next-to-nearest-neighbor (NNN) dimer correlations 
$G^{\,n}_{\alpha}(z,t)=G_{\alpha}(z,t)/\! \int \! dz \, G_{\alpha}(z,t)$, with $\alpha=N\!N,N\!N\!N$ and
\begin{align}
 G_{\scriptscriptstyle \!N\!N}(z,t)= \frac{1}{N_m -1} \sum_m G_{m,m+1}(z,t), \\
 G_{\scriptscriptstyle \!N\!N\!N}(z,t)= \frac{}{N_m -2} \sum_m G_{m,m+2}(z,t),
\end{align}
and we introduce functions 
\begin{eqnarray}
\chi_{\alpha}(t)=G^{\,n}_{\alpha}(0,t)/{\bar G}^{\,n}_{\alpha}(t)
\end{eqnarray}
where ${\bar G}^{\,n}_{\alpha}(t) = \int \! dz \, G^{\,n}_{\alpha}(z,t)^2$ is the mean value of 
$G^{\,n}_{\alpha}(z,t)$. The checkerboard soliton arrangement is characterized by the NN 
anti-correlation ($\chi_{\scriptscriptstyle \,\!N\!N}(t)<1$) and the NNN correlation 
($\chi_{\scriptscriptstyle \,\!N\!N\!N}(t)>1$). In the following we consider a particular case of four 
lattice sites. A generic example of $\chi_{\scriptscriptstyle \,\!N\!N}(t)$ and $\chi_{\scriptscriptstyle \,\!N\!N\!N}(t)$ 
time evolution within the crystalline regime (see discussion below) is 
depicted in the inset of Fig.~\ref{fig:checkerboard}~(top). 

As in the case of filamentation, the emergence of the soliton crystal is limited to a window of $|g_d|/g$ 
values. While for a weak dipolar coupling the system remains stable, a sufficiently large dipole moment 
value renders the attractive intra-tube interactions dominant and, in consequence, we observe the formation 
of the staggered soliton pattern. Note that this configuration, originating in the anti-correlated 
modulational instability emerging within the linear regime, is indeed a highly metastable state, as 
it maximizes NNN dipolar interactions. Crucially, however, our numerical simulations show that such 
soliton crystal state characterized by $\chi_{\scriptscriptstyle \,\!N\!N}(t)<1$ coinciding with 
$\chi_{\scriptscriptstyle \,\!N\!N\!N}(t)>1$ remains stable well beyond typical experimental 
timescales, being hence effectively permanent. Beyond a critical value of the dipolar 
coupling the NNN repulsion destroys the NNN anticorrelation and hence the crystal.

The instability properties of the soliton crystal may be studied by considering 
the average $\chi_{\,\alpha}$ for different time windows, as depicted in 
Fig.~\ref{fig:checkerboard}~(top). For all $g_d$ values within the unstable regime the 
NN anticorrelation function $\chi_{\scriptscriptstyle \,\!N\!N}(t)<1$ remains constant at all 
times. In contrast, depending on the value of the $g_{d}$ parameter, 
$\chi_{\scriptscriptstyle \,\!N\!N\!N}(t)$ function shows two distinctive types of time dependence. 
Namely, while in the window of the crystallization regime $\chi_{\scriptscriptstyle \,\!N\!N\!N}(t)$ 
saturates at a value indicating NNN anticorrelation and so the emergence of a stable soliton crystal. 
Contrastingly, for large dipolar interactions the initially anticorrelated $\chi_{\scriptscriptstyle \,\!N\!N\!N}$, 
which originates in the linear regime, decreases in time indicating destruction of the checkerboard pattern.

Hence, for negative $g_d$ values we identify three distinct regimes depicted in 
Fig.~\ref{fig:checkerboard}~(bottom): (i) a stable regime for small dipole values, 
(ii) an unstable regime intrinsically characterized by the dynamical formation of a 
checkerboard soliton crystal, and (iii) a strong dipolar interactions regime in which 
only nearest neighbor anticorrelation is preserved while the next-to-nearest neighbor 
correlation is lost (soliton liquid). In analogy to the filamentation phenomenon, for 
a fixed $\Delta/l_{\scriptscriptstyle \perp}$ value the regimes boundaries depend solely on 
the $|g_d|/g$ ratio. For $\Delta/l_{\scriptscriptstyle \perp}=6$, the crystalliztion regime 
occurs for $0.40<|g_d|/g<0.47$, which for $^{52}$Cr~($^{164}$Dy) requires 
$7.7<a/a_0<9.1$~($70.0<a/a_0<82.9$).

\section{Summary}

In conclusion, the dipolar inter-site interactions in 
a destabilized dipolar BEC confined in a stack of quasi-1D tubes induce an 
interesting dynamics characterized by the development of a 
correlated modulational instability in the non-overlapping sites. 
For a sufficiently large dipole moment this density modulation seeds 
the spontaneous self-assembly of soliton filaments or a soliton checkerboard crystal, 
depending on the sign of the dipolar interactions. Contrary to 
filaments and crystals of individual molecules, filaments and crystals 
of solitons self-assemble spontaneously merely by simple destabilization 
of the condensate. Moreover, we expect that due to the many-body character 
of the constituent solitons the dipole moment necessary for observing these 
structures may be significantly reduced and that they may be attainable 
with partially polarized polar molecules~\cite{Ni:2008} or highly magnetic atoms, 
paving a promising route towards the first realization of 2D patterns of 
solitons in ultra-cold gases and, to the best of our knowledge, 
in nonlinear optics as well.

\acknowledgements
We acknowledge the support of the Center of Excellence QUEST, the German-Israeli Foundation and the DFG (SA1031/6).

%%%%%%%%%%%%%%%%%%%%%%%%%%% 
%%%%%% BIBLIOGRAPHY: %%%%%%
%%%%%%%%%%%%%%%%%%%%%%%%%%% 
\bibliographystyle{apsrev4-1}
\bibliography{SolitonCrystallization}{}

\end{document}